May 14, 2015

# A New Approach for Scalable Analysis of Microbial Communities

Ehsaneddin Asgari[1], Kiavash Garakani[1] and Mohammad R.K Mofrad[1,2*]

[1] Molecular Cell Biomechanics Laboratory, Departments of Bioengineering and Mechanical Engineering, University of California, Berkeley, CA 94720, USA, [2] Physical Biosciences Division, Lawrence Berkeley National Lab, Berkeley, CA 94720, USA.

**Abstract**
**Motivation:** Microbial communities play important roles in the function and maintenance of various biosystems, ranging from human body to the environment. Current methods for analysis of microbial communities are typically based on taxonomic phylogenetic alignment using 16S rRNA metagenomic or Whole Genome Sequencing data. In typical characterizations of microbial communities, studies deal with billions of micobial sequences, aligning them to a phylogenetic tree. We introduce a new approach for the efficient analysis of microbial communities. Our new reference-free analysis technique is based on n-gram sequence analysis of 16S rRNA data and reduces the processing data size dramatically (by $10^5$ fold), without requiring taxonomic alignment.
**Results:** The proposed approach is applied to characterize phenotypic microbial community differences in different settings. Specifically, we applied this approach in classification of microbial communities across different body sites, characterization of oral microbiomes associated with healthy and diseased individuals, and classification of microbial communities longitudinally during the development of infants. Different dimensionality reduction methods are introduced that offer a more scalable analysis framework, while minimizing the loss in classification accuracies. Among dimensionality reduction techniques, we propose a continuous vector representation for microbial communities, which can widely be used for deep learning applications in microbial informatics.
**Availability:** The Matlab code and data will be available on: http://llp.berkeley.edu.
**Contact:** mofrad@berkeley.edu

## 1 Introduction

It is becoming established that microbial communities have important functions relevant to supporting biological systems. It has recently been shown that microbial life plays a key role in regulating a broad range of systems, ranging from the human body to the soil and water microbiomes in the environment (East, 2013; Chaparro *et al.*, 2012; Pinto *et al.*, 2012). Studies on the characterization of healthy versus unhealthy human microbiomes in different body sites and disease conditions have demonstrated a high potential for microbial community profiling techniques in the diagnosis and treatment of diseases (Marsland *et al.*, 2013; Cho and Blaser, 2012; Mendes *et al.*, 2011). Similarly, classification of the microbiomes has significance in the environment. For instance, the soil microbiome has a great impact on plant fertility(Chaparro *et al.*, 2012), and analyzing microbial communities in drinking water is one of the primary concerns of environmental sciences (Pinto *et al.*, 2012).

The human microbiome is the set of all microorganisms that live in close association with the human body. These communities consist of a variety of microorganisms, including eukaryotes, archaea, bacteria and viruses. Collectively, the number of microorganisms in the human body is approximately ten times more than human cells, and the number of genes are approximately 1,000 times more than that of the human genome. It is now widely believed that changes in the composition of our microbiomes correlate with numerous disease states, raising the possibility that manipulation of these communities may be used to treat disease (Jorth *et al.*, 2014; Cho and Blaser, 2012; Marsland *et al.*, 2013; Mendes *et al.*, 2011; Ramezani and Raj, 2014).

Launched in 2008, the Human Microbiome Project (HMP) was a five-year National Institute of Health (NIH) initiative aimed at determining the role of the human microbiome in the maintenance of health and the onset and progression of disease (Peterson *et al.*, 2009). Using existing metagenomic shotgun sequencing and metagenomic 16S sequencing techniques, extensive information on the role of the human microbiome in health and disease was collected (Hodkinson and Grice, 2015; Langille *et al.*, 2013). The 16S rRNA sequencing is a common amplicon sequencing method used to identify and compare bacteria present in a given sample. The 16S component is a subunit of the 30S RNA present in ribosomes. Due to the slow rates of evolution of this region of the gene, the 16S gene has been especially useful as a taxonomic 'fingerprint' for bacteria. The 16S rRNA gene is composed of nine hypervariable regions (V1-V9) that are used to distinguish bacterial sequences. Sequencing of specific hypervariable regions may be performed for the quantitative determination of microbial community composition.

In this study, we introduce a new scalable approach for the analysis of microbial communities using 16S metagenomic data and demonstrate that this method can effectively distinguish between microbiomes from different body sites, healthy and diseased oral microbiomes, and infant microbiomes over varying times. While existing technologies for microbial analysis require extensive and computationally expensive post-processing for taxonomic binning of 16S sequences, the proposed approach performs no such binning, and relies solely on the 16S sequences of the data inputted. The present method will better incorporate the sequence similarity across bacterial types for the representation and characterization of microbial communities, while being computationally cost effective.

## 2 Methods

### 2.1. The Microbial Words (microWords) Representation

Current technologies for microbial profiling are based on taxonomic phylogenetic alignment using 16S rRNA or Whole Genome Sequencing (WGS) data (Matsen, 2015; Hamady and Knight, 2009; Tanaseichuk *et al.*, 2014; Su *et al.*, 2014). Aside from the computational complexity of phylogenetic alignment, the existing approaches are not capable of incorporating bacteria whose 16S rRNA gene or reference genomes are unknown.

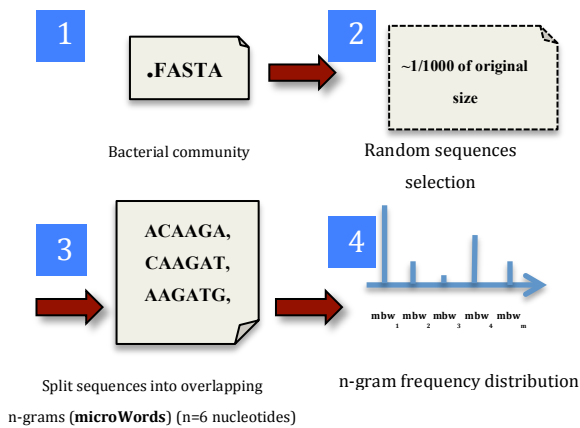

**Figure 1** Key steps involved in feature extraction from each sample. (1) We use bacterial community sequence data in FASTA file, (2) N 16S sequences will be selected randomly from each FASTA file. For consistency reasons, the random subsets are selected only from reads generated using the same dominant primers (e.g. primers targeting the hypervariable regions V3-V5 of the 16S gene), (3) Using an overlapping window of size 6 on the set of N selected sequences, we split each sequence into a number of words 6 letters long (6-grams), called microbial word (microWords), (4) We use frequency distributions of microbial words as the representations of microbial communities in 16S rRNA processing.

Our goal here is to introduce an efficient data analysis method for microbial community analysis. One of the primary steps in each data analysis task is finding an interpretable representation of data for machines that can increase performance of learning algorithms (Bengio *et al.*, 2013). This step, called feature extraction, can influence performance of algorithms significantly. Even the most sophisticated algorithms would perform poorly if inappropriate features are used, while simple methods can potentially perform well when they are fed with the appropriate features (Bengio *et al.*, 2013). In order to extract meaningful features from sequence data that incorporate both bacterial community distribution and sequence similarity, we consider a "bag of words" (Jones, 1972) representation of n-grams (here 6-grams) from a random subset of sequence files (Cavnar *et al.*, 1994; Mantegna *et al.*, 1995). To do so, we randomly select N (here N=1,000) instances from each FASTA file, where a FASTA file is defined as a file obeying the FASTA format, a format that is used for the representation of nucleotide sequences. Importantly, while we selected only 1000 instances from each FASTA file out of the 100,000s that are typically present in each file, the sampling size may be increased according to the needs of the user. Presumably, increasing the sample size of instances would improve the accuracy of representation of low-abundance bacterial species in the sampled communities, but would also result in an increased computational load To keep the consistency of data, we selected the random subset only from reads pertaining to the V3-V5 region of the 16S gene.

Subsequently, using a fixed length-overlapping window on the set of N selected sequences, we split each 16S sequence into a number of words 6 base pairs long (6-grams) called microbial word (**microWords**). The total vocabulary size of 6-grams is $4^6$=4,096 words. Using this method, we present each sampling file of few 100,000s sequences with only a frequency distribution over 4,096 microbial words estimated from word counts, in 1000 random sequences in the file. We use these frequency distributions as the representations of microbial communities in 16S rRNA processing (Figure 1).

Studies in microbial community analysis deal with a large number of samples. Thus, at the end we can present all data with a 2-dimensional matrix (**microMatrix**), in which each row represents a sample file and each column corresponds to the frequency of a **microWord**. Consequently, a set of billions of bacterial sequences can be compactly represented by a real matrix of size of M (*M*: number of samples) by 4,096 (*m*: number of **microWords**):

$$\mu b Matrix = \begin{pmatrix} & \mu bw_1 & \mu bw_2 & \cdots\cdots\cdots & \mu bw_{m-1} & \mu bw_m \\ S_1 & tf_{11} & tf_{12} & \cdots\cdots\cdots & tf_{1(m-1)} & tf_{1m} \\ S_2 & tf_{21} & tf_{22} & \cdots\cdots\cdots & tf_{2(m-1)} & tf_{2m} \\ \vdots & \vdots & \vdots & \cdots & \cdots & \vdots \\ S_{M-1} & tf_{(M-1)1} & tf_{(M-1)2} & \cdots\cdots & tf_{(M-1)(m-1)} & tf_{(M-1)m} \\ S_M & tf_{M1} & tf_{M2} & \cdots\cdots & tf_{M(m-1)} & tf_{Mm} \end{pmatrix}$$

Since the sequences are randomly selected from each sample file, this matrix can be a good representation of the actual sequence composition.

In a typical characterization of microbial communities, due to statistical sufficiency purposes researchers deal with large sets (10s to 1000s) of 16S metagenomic samples collected from different diseased individuals or environments. A typical sample may have about a few hundred thousands of 16S rRNA sequences, resulting in a total size of approximately a billion sequences for analyses. Obviously, processing of such a large amount of data is computationally expensive. Our proposed approach considers a frequency distribution over contiguous sequences of **n** nucleotides (n-grams) in 16S rRNA genes, which will better incorporate the sequence similarities across bacterial types and enhance the representa-

# Big-data Analysis of Microbial Communities

tion of microbial communities. By taking into account the sequence similarity, this approach is capable of incorporating bacteria whose 16S rRNA gene or reference genomes have not been determined, which represent a major portion within human microbial samples and more so in ecological and environmental samples. This is largely due to the inability of many microbes to grow under laboratory environments.

## 2.2. Dimensionality Reduction Techniques for Microbial Words

Our proposed microWords is a low-dimensional representation in comparison with the original 16S rRNA metagenomic data. However, when the number of samples increases, for the ease of computation and memory saving purposes we need to reduce the vocabulary size of microWords (number of columns: *m*). This may be achieved using various dimensionality reduction techniques. Several methods have been developed for dimensionality reduction (Price *et al.*, 2006; Shlens, 2005; Abdi and Williams, 2010; Platzer, 2013; Amir *et al.*, 2013; van der Maaten, 2013). Here, we consider two standard dimensionality reduction techniques **(1)** principle component analysis (PCA) and **(2)** t-stochastic neighbor embedding (t-SNE), as well as our proposed distributed representation for microbial communities, called **(3) Microbe Vector (microVec).** In the next step, we compare these techniques as applied throughout three different community-analysis tasks. This comparison hopefully will help readers to choose the appropriate method for their problems of interest.

**Principle component analysis (PCA)** is one of the most widely used techniques for dimensionality reduction. PCA provides us with the principal directions (*V*'s) in which the data exhibits maximum variation. These principal directions are the eigenvectors of the data covariance matrix and the eigenvalue corresponding to each eigenvector indicate the variance of data in that particular direction. Thus, by a linear transformation we can project data into a low-dimensional space with principal directions as the axes. Despite its widespread use in modern data analysis, PCA has some limitations. Since PCA has covariance matrix estimation and eigenvalue decomposition, the computational complexity of PCA for *M* samples, each represented with *m* features, is of the order $O(m^2M+m^3)$. In addition, in order to have a reliable estimation of the covariance matrix, PCA requires a large number of samples and typically underperforms with limited numbers of data points. Finally, PCA is a linear transformation with several constraints, and thus cannot address the non-linearity in the data(Shlens, 2005; Abdi and Williams, 2010; Price *et al.*, 2006).

**t-Stochastic Neighbor Embedding (t-SNE)** is a dimensionality reduction technique which has been successfully applied to visualization problems, since it attempts to preserve pairwise distance distribution of points in the lower dimensions(Maaten and Hinton, 2008). As the projection in the lower dimensions involves the distribution of relative distances, it requires large amounts of data points to find a meaningful representation. t-SNE is a nonlinear transformation with a computational complexity of $O(M^2)$. However, a recent implementation of t-SNE has been reported to have the computational complexity of $O(M\log(M))$ (van der Maaten, 2013). t-SNE is a relatively new technique in machine learning, and only recently has been used for the analysis of biological data (Platzer, 2013; Amir *et al.*, 2013; McCarthy, 2013; Mnih *et al.*, 2015; Irish, 2014).

**Microbe Vector (microVec)** is a new distributed representation we introduce here to represent the sequence composition of microbial communities. Distributed representation has proved as one of the most successful approaches in machine learning and natural language processing (NLP) (Collobert *et al.*, 2011; Mikolov, Chen, *et al.*, 2013; Mikolov, Corrado, *et al.*, 2013). The main idea in this approach is encoding and storing information associated with a component of a system by establishing its interactions with other members. Distributed representation was originally inspired by the structure of human memory, where the items are stored in a "content-addressable" fashion (Hinton *et al.*, 1986).

In the training process of distributed representation of words in natural languages, a large corpus of sentences should be fed into the training algorithm to ensure sufficient sampling of different contexts. Similarly, a large corpus is needed to train distributed representation for biological sequences. The next step is to break the training sequences (i.e. microbial sentences) into sub sequences (i.e. microbial words). The simplest and most common technique in bioinformatics to study sequences involves fixed-length n-grams (here 6-grams). Previous study of distributed representation for protein sequences suggested multiple lists of shifted non-overlapping words for training (Asgari and Mofrad, 2015). We use the Human Microbiome Reference Genome as a rich microbial sequence database and split it into lists of multiple non-overlapping 6-grams (**microWords**), yielding ~1.5 billion microWords occurrences.

The next step is training the distributed representation based on the prepared data set through a Skip-gram neural network. Using the Skip-gram neural networks, segmented sequences (**microWords**) are represented with a single dense n-dimensional vector with real values (here 100 dimensions). In training such word vector representations, the Skip-gram approach attempts to maximize the probability of observed word sequences (contexts) (Mikolov, Corrado, *et al.*, 2013). In other words, for a given training sequence of words we seek to find their corresponding n-dimensional vectors maximizing the following average log of probability function. Such a constraint allows similar words to assume a similar vector representation.

$$arg\max_{v,v'} \frac{1}{N} \sum_{i=1}^{N} \sum_{-c \leq j \leq c} \log p(w_{i+j}|w_i)$$

$$p(w_{i+j}|w_i) = \frac{\exp(v'^T_{w_{i+j}} v_{w_i})}{\sum_{k=1}^{W} \exp(v'^T_{w_{i+j}} v_{w_i})}$$

where N is the length of the training sequence, 2c is the window size we consider as the context, $w_i$ is the center of the window, W is the vocabulary size and $v_w$ and $v'_w$ are input and output n-dimensional representations of word w, respectively. The probability $p(w_{i+j}|w_i)$ is defined using a softmax function (Mikolov, Chen, *et al.*, 2013). In the implementation of our proposed method, we use (Word2Vec) (Mikolov, Corrado, *et al.*, 2013), which is considered as the state-of-the-art for training word vector representation (Goldberg and Levy, 2014). To train the embedding vectors, we consider a vector size of 100 and a context size of 20. Thus, each **microWords** is presented as a vector of size 100, termed microbe-vector (**microVec**), and each sequence is represented as the summation of the vector representation of microVecs. Subsequently, to represent the microbial community we multiply the microMatrix$_{(M \times m)}$ by the matrix of all microVecs$_{(m \times 100)}$. Therefore, the whole microbial communities in all samples can be presented by a matrix of size M×100.

Microbe-vector needs to be trained only once based on a large collection of sequences, and can then be used to ascertain a diverse set of information regarding the microbial communities of interest. Since the distributed vector representation learning can be conducted offline, the computational complexity of finding this transformation can be done in $O(1)$, i.e. substantially faster than t-SNE. Since eigenvectors of PCA can also be calculated offline, in Table 1 the complexity of PCA is mentioned to be either $O(m^2M+m^3)$ or O(1). Another advantage of **microVec** (as well as offline PCA) over t-SNE is associated with data-size invariance. Since **microVec** is pre-trained over a large number of sequences and has encoded a language model of microbial sequences in an offline manner, the representation of each microbial sample doesn't change with the number of samples. Furthermore, skip-gram neural network has a nonlinear architecture. Thus, **microVec** representation is also a nonlinear transformation of data. Table 1 shows a brief comparison of PCA, t-SNE, and microVec. The microVec representation can also be considered as pre-training for various applications of deep learning on microbial sequence data (Asgari and Mofrad, 2015; Alipanahi *et al.*, 2015; Zhou and Troyanskaya, 2015; Park and Kellis, 2015), which is a state-of-the-art approach in bioinformatics and more generally in machine learning (Dahl *et al.*, 2012; Erhan *et al.*, 2010).

**Table 1 Comparison of different dimensionality reduction methods for microbial communities.**

|  | Computational Complexity | Linearity | Data-size invariance |
|---|---|---|---|
| PCA | $O(m^2M+m^3)$ or $O(1)$ | Linear | No |
| t-SNE | $O(M \log M)$ | Nonlinear | No |
| microVec | $O(1)$ | Nonlinear | Yes |

### 2.3. Validation and Application of the Proposed Approach

To assess the ability of the proposed method to represent microbial communities, we performed classification of the data based on phenotypic microbial community differences. Classification in machine learning is a task of assigning categories to data points whose categorical information is unknown, based on a set of training samples with known categorical information. In classification, a mathematical function, called a classifier, is produced that can predict the category of each new data point (test data) based on training examples (Suykens and Vandewalle, 1999).

In order to evaluate a classifier, we often split the data points with known categories into two disjoint sets: training set and testing set. Subsequently, the classifier will learn from the training set and its performance will be evaluated against the test set. In order to gain insight on how the classifier will perform on an independent new dataset, we often try different training and test sets. 10xFold cross-validation refers to dividing the data points with known categorical information into 10 disjoint folds and performing the training on 9 of them and evaluating the trained classifier against the remaining fold. Since we can perform the training/evaluation in 10 different ways, eventually we report the average and standard deviation for 10 accuracies of 10 different ways (Refaeilzadeh *et al.*, 2008). Several classification methods exist in machine learning (e.g., support vector machine, Gaussian, decision tree, random forest, etc.) (Suykens and Vandewalle, 1999; Breiman, 2001; Yau and Manry, 1990; Safavian and Landgrebe, 1991). However, data representation contributes the most to the accuracy of classification. Choosing a relevant representation would result in a higher accuracy. Thus, in order to prove the relevance of our proposed representation for microbial community analysis, we evaluate it throughout classification tasks.

We perform three different tasks: 1) classification between microbial communities that belong to different/multiple body sites, 2) classification between oral microbiomes associated with healthy and diseased individuals, and 3) prediction of microbial communities over time during the development of infants.

#### 2.3.1. Characterization of body sites compositional differences

We employ the metagenomic 16S sequence dataset provided by the NIH Human Microbiome Project (Peterson *et al.*, 2009). In particular, we use processed, annotated 16S sequences of up to 300 healthy individuals, each sampled at 4 major body sites (oral, airways, gut, vagina) and up to three time points. For each major body site, a number of subsites were sampled. We focused on 5 body subsites: anterior nares (nasal) with 295 samples, saliva (oral) with 299 samples, stool (gut) with 325 samples, posterior fornix (urogenital) with 136 samples, and mid vagina (urogenital) with 137 samples. These body sites are selected to represent differing levels of spatial and biological proximity to one another, based on relevance to pertinent human health conditions potentially influenced by the human microbiome.

Majority of body sites are sampled at hypervariable regions V3-V5 of the 16S gene. Preliminary studies have examined the efficacy of different hypervariable regions within the 16S rRNA for taxonomic classification purposes (Chakravorty *et al.*, 2007), yet based on availability and current standards for the analysis of 16S sequencing we choose to exclusively use the V3-V5 region (although this method may be used for the analysis of other hypervariable regions). Thus, for consistency purposes we filter out sequences from different primers. Specifically, we use 258 samples from anterior nares (nasal), 278 samples from saliva (oral), 312 samples from stool (gut), 130 samples from posterior fornix (urogenital), and 130 samples from mid vagina (urogenital), in total 1,108 samples.

As mentioned in **Error! Reference source not found.**, from each sample file we select 1,000 random sequences and present each file with microWords distribution of its 1,000 randomly selected sequences, hence we have 1,108 microWords distribution corresponding to sample files. As the next step, we evaluate how this representation can distinguish between phenotypic differences. For this purpose, we use a multi-class support vector machine classifier throughout a 10xfold cross validation task (Suykens and Vandewalle, 1999). A high accuracy of supervised classification will show us how phenotypic differences can be discriminated using our approach. Furthermore, to address big-data processing we perform the same task after performing dimensionality reduction methods: PCA, t-SNE, and microVec to compare their performance.

#### 2.3.2. Differentiation of healthy and diseased oral human microbiome

We use the data provided by (Jorth *et al.*, 2014) to determine the ability of our approach to differentiate between healthy and diseased periodontal microbiomes. . In (Jorth *et al.*, 2014), microbial samples were collected from subgingival plaques from 10 healthy patients and 10 patients with periodontitis. This data included RNA sequencing data of 16S rRNA sequences using primers that amplify the V4/V5 regions of the 16S rRNA. Here, each sample is subsequently subsampled randomly to obtain a larger data set for visualization and classification. We perform random sequence selection from each individual sample file 100 times. This helps us to see the intra- and inter-file differences. Each sampling is represented with microWords distributions. In order to distinguish between healthy and diseased periodontal microbiomes, we use binary-class support vector machine classifier throughout a 10xfold cross validation task. The level of accuracy of supervised classification would indicate whether healthy and disease cases could be discriminated using our approach. Furthermore, we conduct the same classification task after performing dimensionality reduction methods: PCA, t-SNE, and microVec.

#### 2.3.3. Differentiation of microbiome composition over infant development

The data provided by (Koenig *et al.*, 2011) is used to assess the ability of our method to differentiate between microbial communities over differing times during infant development. In the data, 63 16S rRNA sequences of stool samples in a single healthy infant are collected over a 2.5 years time period. We perform subsampling of each sample to obtain a larger data set for testing. Random sequence selection is performed on each individual sample file 100 times. This helps us to see the within and between file differences. Each sampling is represented with microWords distributions. In the next step, we again perform the same classification task after applying dimensionality reduction methods: PCA, t-SNE, and microVec.

## 3  Results

### 3.1. Characterization of body sites compositional differences

In order to determine whether the composition of 16S microbial sequences is sufficient for classification of varying body sites, we performed t-SNE on the 6-gram distributions of each sample (shown graphically in Figure 3). Samples formed four well-defined clusters. Interestingly, our results revealed that most of the clustered samples were from the same major body site (nasal, oral, gut, or urogenital), suggesting that the composition of microbial sequences is sufficient to distinguish the major body sites (nasal, oral, gut, and urogenital).

Next, we attempted to classify body sites based on 16S sequence information. Using a SVM classifier on our microWords distribution, we generated the confusion matrices for the 4-class (classification between nasal, oral, gut, and urogenital sites) classification problem (Table 2). Shown in Error! Reference source not found., SVM classification on the microWords distribution demonstrated highly accurate prediction of major body sites from 16S sequences and we obtained a classification accuracy of 98.60% ± 1.78% (mean ± s.d.).

To ensure reasonable scalability of our method, we tested the classification accuracy of various dimensionality reduction methods. As shown in Table 3, classification using PCA, t-SNE, and microVec yielded classification accuracies of 98.40% ± 1.07%, 98.20% ± 1.92%, and 96.20% ± 1.93%, respectively.

**Table 2 Confusion matrix for 4-class classification**

|  | Nasal | Urogenital | Oral | Gut |
|---|---|---|---|---|
| **Nasal** | 248 | 1 | 5 | 4 |
| **Urogenital** | 1 | 164 | 0 | 0 |
| **Oral** | 0 | 0 | 278 | 0 |
| **Gut** | 3 | 0 | 0 | 309 |

Minor differences in classification accuracy of body sites using dimensionality reductions on microWords distribution indicated that when a minor error is allowed, we can apply a proper dimensionality reduction technique to reduce computational complexity. Furthermore, although microVec had lower classification accuracy than those of PCA and t-SNE, it was significantly faster (**Table 1**).

**Table 3 Comparison of the performance of 6-grams in body site classification with different dimensionality reduction methods.**

| Classification Method | Dimensionality | Classification Accuracy |
|---|---|---|
| microWords distribution | 4,096 | 98.60% ± 1.78% |
| PCA | 100 | 98.40% ± 1.07% |
| t-SNE | 100 | 98.20% ± 1.92% |
| microVec | 100 | 96.20% ± 1.93% |

### 3.2  Differentiation of healthy and diseased oral human microbiome

Next, we tested the ability of our method to distinguish healthy and diseased oral microbiomes in periodontal disease. In the provided data, the oral microbiomes of 10 healthy and 10 diseased individuals were sampled. We performed t-SNE on the microWords distributions of each subsample (Figure 2). As can be seen in Figure 2, samples clustered very well according to individual, forming a total of 20 well-defined clusters. Furthermore, samples from different individuals qualitatively clustered according to the health state of the microbiome.

To quantitatively assess the ability of our method to classify healthy versus diseased microbiomes, we performed a classification of healthy versus diseased microbiomes. Using this, we obtained a classification accuracy of 100% ± 0.00%. Furthermore, application of dimensionality reduction techniques for scalable analysis, including PCA, t-SNE, and microVec resulted in no reduction in classification accuracy (Table 1).

**Table 4- Comparison of the performance of microWords distribution in disease condition classification of bacterial community with different dimensionality reduction methods.**

| Classification Method | Dimensionality | Classification Accuracy |
|---|---|---|
| microWords distribution | 4,096 | 100% ± 0.00% |
| PCA | 100 | 100% ± 0.00% |
| t-SNE | 100 | 100% ± 0.00% |
| microVec | 100 | 100% ± 0.00% |

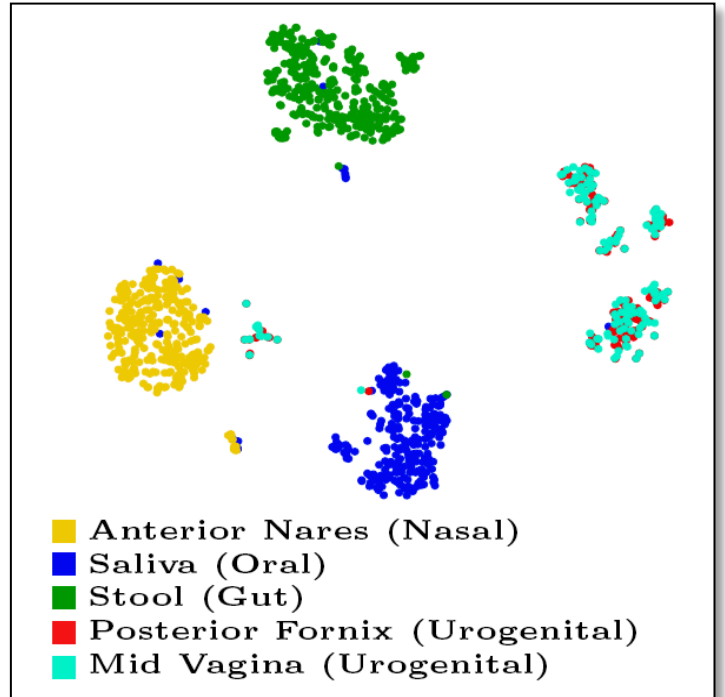

**Figure 3** Visualization of samples using t-SNE on 6-gram data. Samples are colored according to body sites of sampling.

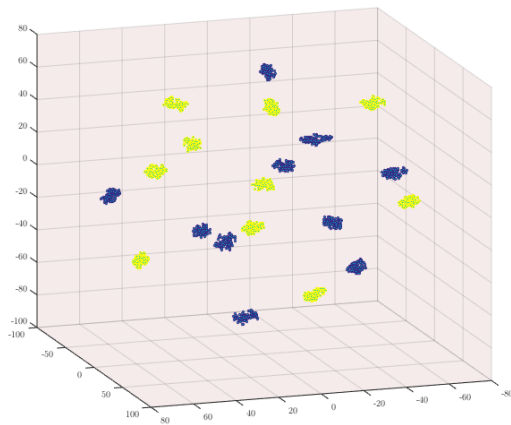

**Figure 2** Visualization of disease vs. healthy microbiomes using t-SNE on the 6-gram data. Blue and yellow dots represent disease and healthy samples, respectively. Samples from the same individual are clustered together

### 3.3 Differentiation of microbiome composition over infant development

Our community analysis technique was next used for the classification of microbiome populations according to infant's age. As shown in Table 5 Comparison of the performance of 6-grams in age classification from bacterial community with different dimensionality reduction methods., microbiome samples were classified according to corresponding infant age with 99.95% ± 0.00%, with only minimal reduction in classification accuracy upon application of more scalable dimensionality reduction techniques.

**Table 5** Comparison of the performance of 6-grams in age classification from bacterial community with different dimensionality reduction methods.

| Classification Method | Dimensionality | Classification Accuracy |
|---|---|---|
| microWords distribution | 4,096 | 99.95% ± 1.90% |
| PCA | 100 | 99.90% ± 0.36% |
| t-SNE | 100 | 99.90% ± 0.36% |
| microVec | 100 | 98.50% ± 3.17% |

## 4 Conclusion

A new scalable approach was introduced for the analysis of microbial communities. Using only 16S sequence data of microbial samples, the proposed method was employed to accurately distinguish microbial communities based on phenotypic differences. In particular, we obtained high classification accuracies in categorizing the data according to body sites, differentiation of periodontal microbial communities in health and periodontitis as well as characterization of microbial samples taken at different times in developing instants. We also suggested different dimensionality reduction methods that could provide us with a more scalable analysis framework, while having a minor loss in classi-

# Big-data Analysis of Microbial Communities

fication accuracy. While highly computationally efficient, this approach does not require taxonomic alignment and will better incorporate the sequence similarities across bacterial types and enhance representation and characterization of microbial communities. By taking into account the sequence similarity, this approach is capable of incorporating bacteria whose 16S rRNA gene or reference genome have not been determined, which represents a major portion within human microbiome samples and even more so in ecological and environmental samples. We also introduced a vector representation for microbial communities using neural networks. This representation can be utilized as pre-training for deep learning on microbial sequence data.

## ACKNOWLEDGEMENT

Fruitful discussions with Ali Madani and Hengameh Shams are gratefully acknowledged.